\documentclass[aps,prd,amsfonts,preprint]{revtex4}
\usepackage[dvips]{graphicx}
\usepackage{bm}

\begin{document}

\title{Primordial Bispectrum Information from CMB Polarization}

\author{Daniel Babich}

\email{babich@physics.harvard.edu}

\author{Matias Zaldarriaga}

\email{mzaldarriaga@cfa.harvard.edu}

\affiliation{Deparment of Physics, Harvard University, MA 02138}

\affiliation{Harvard-Smithsonian Center for Astrophysics, 60 Garden Street, Cambridge, MA 02138}

\date{\today. To be submitted to Phys. Rev. D.}

   \begin{abstract}
	After the precise observations of the Cosmic Microwave Background (CMB) anisotropy power spectrum,
	attention is now being focused on higher order statistics of the CMB anisotropies. Since linear 
        evolution  preserves the statistical properties of the initial conditions, observed non-Gaussianity
	of the CMB will mirror primordial non-Gaussianity. Single field slow-roll inflation robustly predicts 
        negligible non-Gaussianity so an indication of primordial non-Gaussianity will suggest alternative 
        scenarios need to be considered. In this paper we calculate the information on primordial 
        non-Gaussianities encoded in the polarization of the CMB. 
	After deriving the optimal weights for a cubic estimator we evaluate the Signal-to-Noise 
	ratio $(S/N)$ of the estimator for WMAP, Planck and an ideal cosmic variance limited experiment. 
        We find that when the experiment
  	can observe CMB polarization with good sensitivity, the sensitivity to primordial non-Gaussianity 
        increases by roughly a factor of two. We also test the weakly non-Gaussian assumption used to
	derive the optimal weight factor by calculating the degradation factor produced by the 
        gravitational lensing induced connected four-point function. 
	The physical scales in the radiative transfer functions 
        are largely irrelevant for the constraints on the primordial non-Gaussianity. 
        We show that the total $(S/N)^2$ is simply proportional to the number of observed pixels on the sky.  	
   \end{abstract}

\maketitle

\section{Introduction}

Recent advances in observational cosmology have led to unprecedented constraints on the parameters of the cosmological model. One of the outstanding goals in the field is to determine the mechanism responsible for the seeds that lead to the the structure in the observable Universe. The current data is consistent with an initial scale invariant and adiabatic spectrum of primordial curvature perturbations which existed outside the horizon at recombination. These observations are in agreement  with the predictions of single field slow-roll  inflationary models, but various alternatives scenarios still remain viable. 

In the next decade we will see further advances in observations that will constrain many aspects of the primordial seeds.  
The spectral index of their power spectrum will be measured more accurately and over a wider range of scales by a 
combination of Cosmic Microwave Background (CMB) and other Large Scale Structure (LSS) probes. Dedicated CMB 
polarization instruments will establish  whether or not there is a stochastic background of gravitational waves as 
predicted by models where inflation happens at the GUT scale.  The constraints on Gaussianity, which are the focus 
of this paper, will also improve significantly.  

The Gaussianity of the primordial fluctuations has received a lot of attention lately mainly because in standard 
single-field slow-roll inflation models the deviations from Gaussianity of the perturbations can be fully calculated and are directly related to the departures from scale invariance and thus predicted to be very small \cite{maldacena}, most probably unobservable in the CMB. Thus constraints on the Gaussianity can help distinguish simple inflation models from the various alternatives. Most of the alternatives to single-field slow-roll inflation solve the standard cosmological problems in the same way, by invoking a period of accelerated expansion  in the very early universe (see \cite{cyclic} for a counter example). They differ however in the characteristics of the produced perturbations. Just as  in slow-roll models the perturbations arise from quantum fluctuations during inflation but the detailed physics is not the same. In some of the models the dynamics of the inflaton field is fundamentally changed by the presence of higher derivative terms in the Lagrangian which can even lead to an inflaton field that is not rolling slowly \cite{mukh,creminelli,nima,dbi}. Another possibility are models in which fluctuations in another field different from the inflaton are responsible for the adiabatic fluctuations we observe today \cite{decay,curvaton}. 

These alternative models usually make distinctive predictions about the shape of the spectrum of primordial 
perturbations, the amplitude of the gravity wave background and the departures from Gaussianity as described, 
for example, by the 3-point function. For scale invariant perturbations the three point function in Fourier 
space or bispectrum, is effectively a full function of two variables, thus it could contain a wealth of 
information about the primordial seeds. The different alternatives to slow-roll inflation not only predict 
different levels of non-Gaussianities but also different shapes for the bispectrum as a function of triangle 
configuration. Moreover one can make a very definite and model independent statement about the shape of 
the three-point function in the so called collapsed limit. The collapsed limit corresponds to a three-point 
function where one of the Fourier modes has a much longer wavelength than the other two. In that limit 
the three-point function should go to zero, unless more than one degree of freedom is dynamically relevant 
during inflation \cite{Creminelli:2004yq}. Other consistency relations can be obtained involving the 
three-point function and the amplitude and slope of tensor perturbations \cite{Gruzinov:2004jx}.  

 

Thus a detailed measurement of the three-point function could provide very interesting  information on the mechanism responsible for generating the primordial curvature perturbations. In this work we will use the so called ``Local Model'' for the 
non-Gaussianities and specify the amplitude of non-Gaussianity by the
parameter $f_{NL}$. In this model the gravitational potential, $\Phi(\bm{x})$, can be expressed in terms
of a Gaussian gravitational potential, $\Phi_g(\bm{x})$, as 
\begin{equation}\label{fnl}
   \Phi(\bm{x}) = \Phi_g(\bm{x})+f_{NL}[\Phi_g^2(\bm{x}) - \langle \Phi_g^2(\bm{x}) \rangle].
\end{equation} 
The bispectrum in this model can be written as
\begin{equation}\label{local}
   \langle \Phi(\bm{k}_1) \Phi(\bm{k}_2) \Phi(\bm{k}_3) \rangle = 2f_{NL}(2\pi)^3\delta^{(3)}(\bm{k}_1+\bm{k}_2+\bm{k}_3) 
	[P(k_1)P(k_2) + cyc.], 
\end{equation} 
where $P(k)$ is the power spectrum. 

In general the primordial bispectrum cannot be written as in Eq. (\ref{local}) and so  
constraints on the $f_{NL}$ parameter do not automatically apply to all models of primordial non-Gaussianity \cite{babich1}. 
Nevertheless, we can list approximate values of $f_{NL}$ that are expected in the various alternatives to single field slow-roll inflation 
in order to estimate the detectability of these various models. For comparison in slow-roll inflation models  one expects 
an  $f_{NL} \sim 0.05$ \cite{maldacena}. In models where higher derivative operators are important for the dynamics of the field one can expect results ranging from $f_{NL} \sim 0.1$ in models where high derivative operators are suppressed by a low UV cut-off  \cite{creminelli} to $f_{NL}\sim 100$ in models based on the Dirac-Born-Infeld effective action.  A recent idea called ghost inflation, where inflation occurs in a background that has a constant rate of change instead of a constant background value, can also give $f_{NL} \sim 100$ \cite{nima}. 
The effect of additional light fields on the efficiency of reheating can lead to inhomogeneities in the thermalized
species \cite{decay}; this mechanism was shown to produce an $f_{NL} \ge 5$ \cite{decayNG}.
Another example is the curvaton model where isocurvature fluctuations in a second light scalar field during 
inflation generate adiabatic fluctuations after the inflationary epoch is completed, 
these can cause large non-Gaussianity, $f_{NL} \ge 10$ \cite{curvaton}.

There are two additional sources of non-Gaussianity that although not primordial in origin could be observed first by future experiments. The first and perhaps the most important for observations is secondary anisotropies such as gravitational lensing, the thermal and kinetic Sunyaev-Zeldovich effects and the effects of a patchy reionization. The second source are non-Gaussianities related to the non-linear nature of General Relativity. 
The expectation is that this latter source will lead to $f_{NL} \sim 1$ (eg. \cite{carroll,bartolo}), although there is no full calculation of the temperature  and polarization anisotropies of the CMB to second order in the the curvature perturbations. In  \cite{cremzal} results were presented for the collapsed limit, where the calculation simplifies enormously. 
It that case $f_{NL} \sim 0.7$ was obtained although the shape dependance of the bispectrum was different from 
that of the $f_{NL}$ model. 

While some of the afforementioned models of the early universe are speculative, 
the second order single field, slow-roll inflation and non-linear radiative transfer calculations should 
be taken as definite predictions of standard cosmology. It is interesting to 
determine how close we are to observing any of these effects. Currently the best constraints, which come from
the Wilkinson Microwave Anisotropy Probe (WMAP) \cite{wmapurl}, imply $-58 < f_{NL} < 134$ ($95 \%$ C.L.) 
\cite{wmapfnl}.

The theoretical ability of CMB temperature maps to constrain $f_{NL}$ has been determined for a COBE normalized 
Einstein-de Sitter (EdS) model \cite{komatsu1}. It was shown that the minimum $f_{NL}$ detectable by WMAP is 20, Planck is 5 and an
ideal experiment (no noise and infinitesimal beam width) is 3. The limiting factor in the 
case of an ideal experiment was taken to be the effect of gravitational lensing, which increases the 
estimator noise without affecting the CMB bispectrum signal. Gravitational lensing adds to 
the cosmic variance portion of the noise because the particular realization of the large 
scale structure that lenses the CMB is also {\it a priori} unknown. Other secondary sources of  anisotropies will have a similar effect. 

Since it appears that we are close to being able to detect some of the interesting non-Gaussian effects
mentioned above, it is important to explore other sources of information  beyond the CMB temperature fluctuations to 
see if the minimum detectable value is lowered when the new information is included. The CMB is linearly polarized and 
the $E$ type polarization is sensitive to the primordial curvature fluctuations. 
Therefore in this paper we will ask how much stronger would the constraints on $f_{NL}$ 
be if information from the $E$ polarization is included. Since $B$ polarization cannot be directly 
generated by the scalar primordial curvature fluctuations we will ignore it in this paper.

As we mentioned above, there are two ways in which secondary anisotropies can complicate and degrade our ability to constrain
primordial non-Gaussianity. If the source of secondary anisotropies being considered  produces a three-point function it might bias
the estimator of primordial non-Gaussianity. In practice, a model of primordial non-Gaussianity is assumed
and a reduced bispectrum template is calculated. The distinct shape of the primordial reduced bispectrum
may allow it to be distinguished from the other bispectra \cite{komatsu1}, assuming it can be detected with a sufficiently high 
signal to noise ratio. In addition secondaries add fluctuations which increase the variance estimators of the three-point function 
without contributing to the signal. In fact the variance of the estimator is related to the six-point function of the temperature field. 
The six-point function can be expressed in terms of its unconnected Gaussian contribution: 
permutations of three two-point functions, the product of a two and a connected four-point functions and a connected six-point function. 
Previous  work on this subject has ignored any non-Gaussian contribution to the six-point function. 

\begin{figure}
     \centering
     \includegraphics[width = 7.5cm, height = 7.5cm]{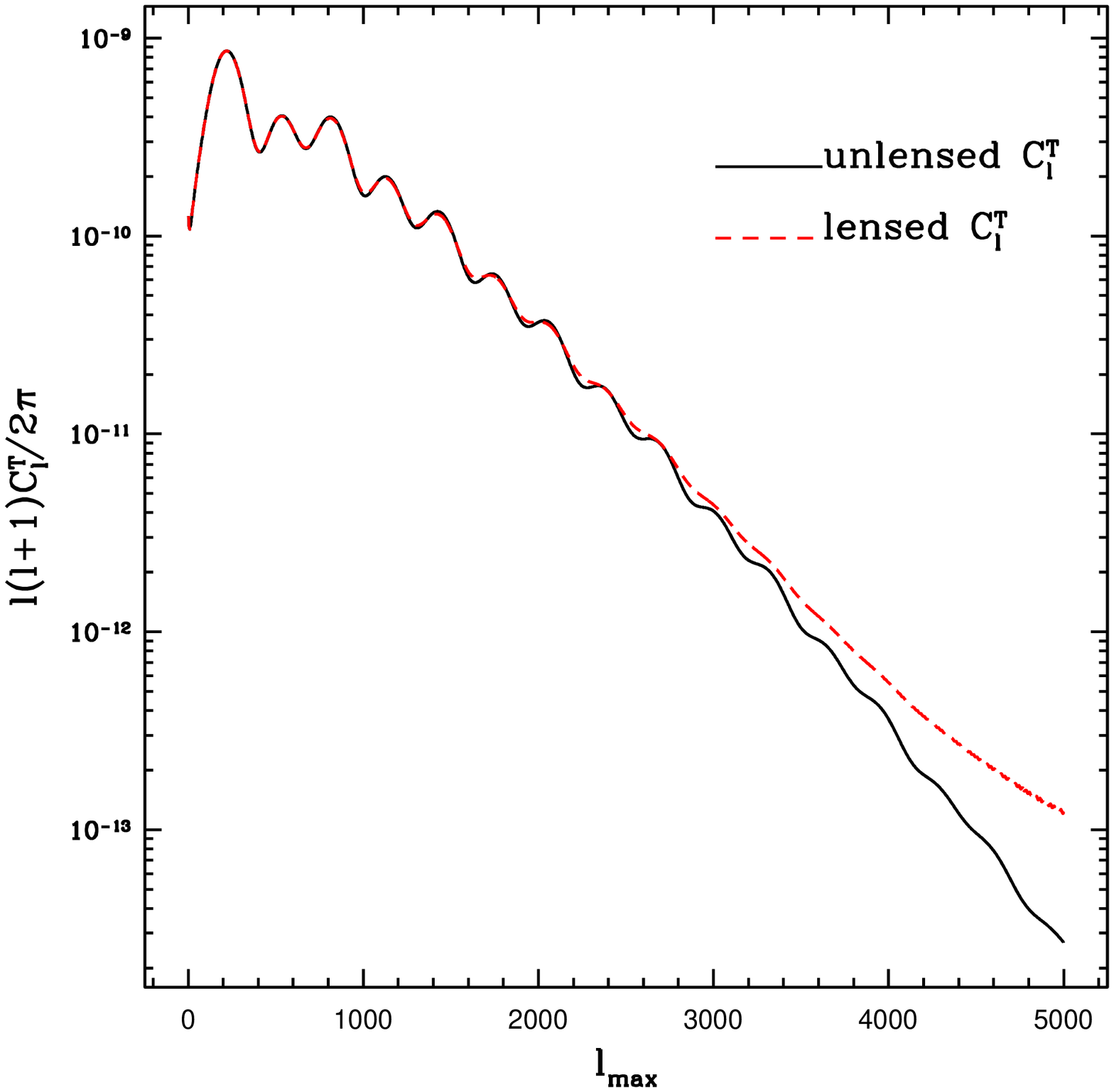}%
     \hspace{5mm}
     \includegraphics[width = 7.5cm, height = 7.5cm]{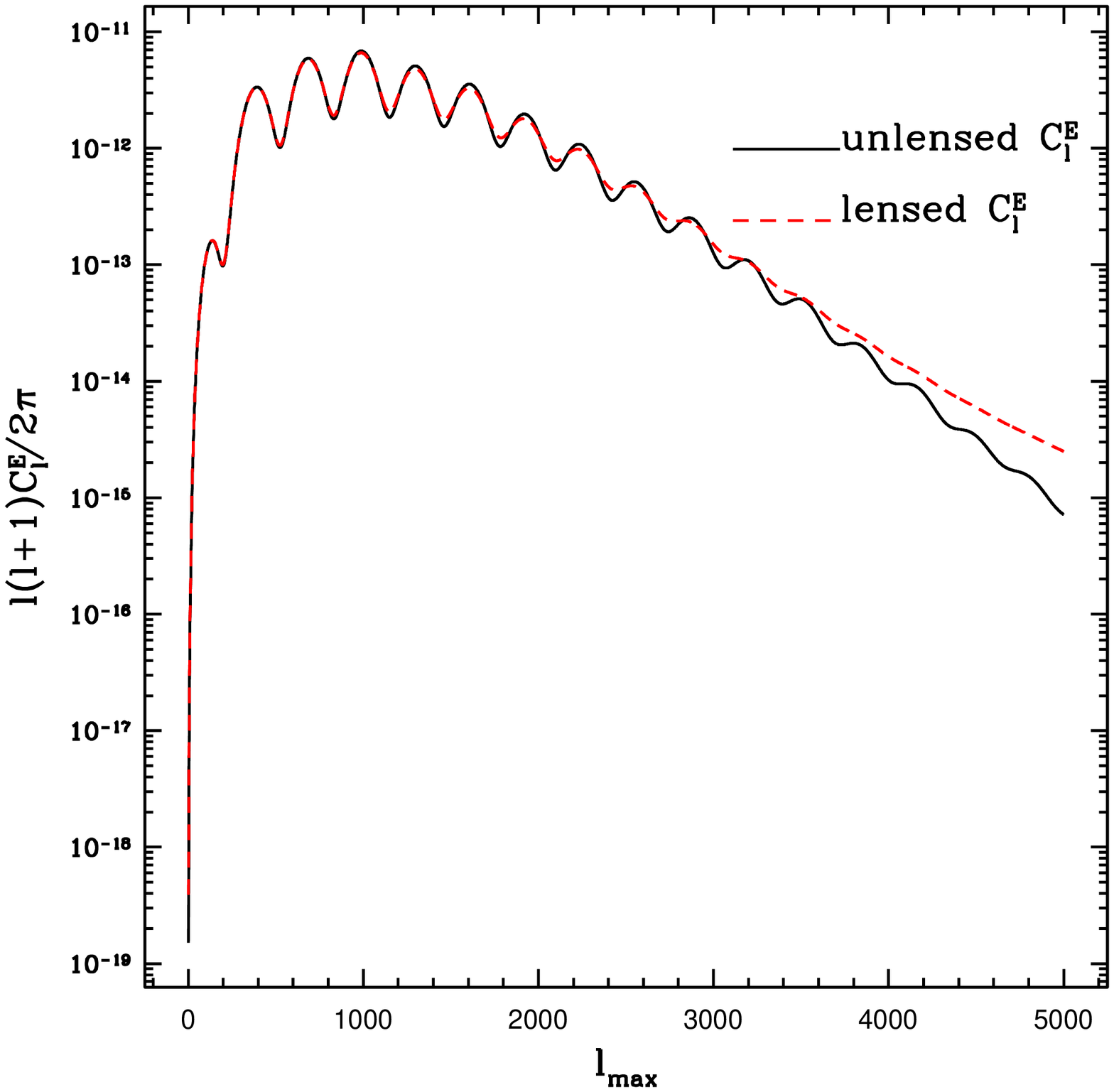}%
     \caption{\label{cl} The lensed (dashed red) and unlensed (solid black) $C_l$ for the 
      concordance $\Lambda CDM$ cosmology; the temperature power spectra are on the left 
      and the $E$ polarization power spectra are on the right.}
   \end{figure}

The secondary anisotropy that will be considered in detail is gravitational lensing. 
Gravitational lensing does not produces a three-point function, so it will not bias our estimator of the
primordial bispectrum. However it will produce corrections to the two-point function and create
four and six-point functions even if the CMB anisotropies are perfectly Gaussian \cite{MZ,Hu2}.
Also, lensing will create small scale power in the CMB two-point function which acts like 
noise in the analysis of primordial non-Gaussianity. All information about $f_{NL}$ is eliminated 
on scales where this effect dominates; this limits the minimum $f_{NL}$ that an ideal experiment 
can detect to be $f_{NL} \sim 3$ for the EdS model \cite{komatsu1}.
Fortunately the graviational lensing effect in a EdS cosmology  is much larger 
than the gravitational lensing effect in a $\Lambda CDM$ concordance cosmology \cite{seljak}, so
 gravitational lensing will be less important in our calculations. For reference
the lensed and unlensed CMB power spectra are plotted in Fig. \ref{cl}. Lensing increases the power
 in the temperature (polarization) anisotropies by a factor of two at $l\sim 4200$ ($l\sim 5000$). 
 However we will find that the four-point function starts reducing the signal to noise of the bispectrum on significantly larger scales. 

In this work we will ignore other secondary anisotropies such as the thermal and kinetic 
Sunyaev-Zeldovich effects and
the Ostriker-Vishniac effect. The dominant secondary anisotropy, the thermal Sunyaev-Zeldovich
effect, has a characteristic spectral shape that will allow it to be separated. In general the
physics of these secondary anisotropies requires a non-linear analysis using hydrodynamical simulations,
we will leave a detailed analysis of the effects of these secondary anisotropies to a future work \cite{zahn}.
These secondary anisotropies also produce polarization in the CMB, fortunately
the amplitude of the polarization secondary anisotropies is much lower than the 
corresponding temperature anisotropies and will be less of a problem.

Finally, although the $E$ and $B$ polarization modes most directly correspond to the primordial curvature fluctuations
and gravity waves, they are not directly measured in CMB experiments. The Stokes' parameters $\mathbf{Q}$ and 
$\mathbf{U}$ are measured and then decomposed into $E$ and $B$. While this 
decomposition is perfectly well defined for a noise free experiment observing the full sky, ambiguities
arises in practical experimental situations \cite{bunn1}. Fortunately the majority of the information on $f_{NL}$
comes from small scales, where this ambiguity is less of a problem if we assume the beam is oversampled
in order to reduce the effects of power aliasing that can also mix $E$ and $B$ modes \cite{antony,bunn2}. 
In this paper we will ignore these complications in the decomposition of the Stokes' parameters.

The paper is organized as follows: in section II we derive the optimal estimator when polarization information is included in the
non-Gaussianity measurements, provide numerical results for the improved constraints 
on $f_{NL}$ and quantify the reduction in the $S/N$ due to the connected four-point function 
contribution to the noise. 
In section III we analytically reproduce our calculations for toy models that do not include the effect
of radiative transfer or the curvature of the sky, but develop intuition about our numerical results.  
We will conclude in section IV with a discussion of our results. In this paper we assume the 
standard $\Lambda CDM$ cosmology with $\Omega_v = 0.73$, $\Omega_b = 0.044$, $\Omega_c = 0.226$,
$\sigma_8 = 0.8$ and $H_0 = 72$ km $s^{-1}$ Mpc$^{-1}$ with a scale invariant primordial power 
spectrum and normalized to the 1-yr WMAP data \cite{wmapdata,wmapurl}. Also we do not include 
the effects of reionization nor the late-time integrated Sachs-Wolfe effect. The late-time
integrated Sachs-Wolfe effect only affects large scales which do not contribute much to the
total signal.

\section{Polarization} 

\subsection{Optimal Estimator}
The Signal-to-Noise ratio $(S/N)$ defined in \cite{komatsu1} can be generalized to include polarization 
information by finding the optimal weight functions for a cubic estimator.
First we form the estimator of the CMB temperature and polarization bispectrum signal as
\begin{equation}
   \hat{S} = \sum_{i,j,k} \sum_{ \{l,m \}}  W^{i,j,k}_{\{l,m\}} a^i_{l_1 m_1} 
     a^j_{l_2 m_2} a^k_{l_3 m_3}, 
\end{equation}
where the indicies $i,j,k$ run over $T$ and $E$, $\{ l,m\}$ indicates all three $l_i,m_i$ 
and $W^{i,j,k}_{\{ l,m \}}$ is the weight function we will  optimize. 
The expectation value of the estimator is 
\begin{equation}\label{reducedbispect}
   \langle S \rangle = \sum_{i,j,k} \sum_{\{ l,m \}}  W^{i,j,k}_{\{l,m\}} 
    \mathcal{G}^{m_1 m_2 m_3}_{l_1 l_2 l_3} b^{i,j,k}_{l_1 l_2 l_3}, 
\end{equation}
where following the standard notation in the literature we separate $\langle a_{l_1 m_1} a_{l_2 m_2} a_{l_3 m_3} \rangle$ 
into the reduced bispectrum $b_{l_1 l_2 l_3}$ and the Gaunt Integral, 
\begin{equation}
   \mathcal{G}^{m_1 m_2 m_3}_{l_1 l_2 l_3} = \sqrt{\frac{(2l_1+1)(2l_2+1)(2l_3+1)}{4\pi}} 
   \left(\begin{array}{ccc} l_1 & l_2 & l_3 \\ 0 & 0 & 0 \end{array}\right)
   \left(\begin{array}{ccc} l_1 & l_2 & l_3 \\ m_1 & m_2 & m_3 \end{array}\right),
\end{equation}
which characterizes the angular dependece of the bispectrum.
The sum over $i,j,k$ includes all eight possible bispectra $\{ TTT,TTE,TET,ETT,TEE,ETE,EET,EEE \}$. 
In the weakly non-Gaussian limit we neglect the contribution of the primordial bispectrum to 
the variance of $\hat{S}$, which becomes
\begin{equation}\label{noise}
   \langle N^2 \rangle = \sum_{i,j,k} \sum_{p,q,r} \sum_{\{ l,m \}} \sum_{\{ t,s \}} 
     W^{i,j,k}_{\{l,m\}} W^{p,q,r}_{\{t,s\}} [\mathbf{Cov}]^{i,j,k|p,q,r}_{\{ l,m|t,s \}}, 
\end{equation}
here we have defined a covariance matrix between the eight possible bispectra for each value of $\{ l,m \}$ 
and $\{ t,s \}$ as 
\begin{equation}\label{cosmicvariance}
   [\mathbf{Cov}]^{i,j,k|p,q,r}_{\{ l,m|t,s \}}  
      = \langle a^{(i)}_{l_1 m_1} a^{(j)}_{l_2 m_2} a^{(k)}_{l_3 m_3} a^{(p)}_{t_1 s_1} a^{(q)}_{t_2 s_2} 
         a^{(r)}_{t_3 s_3} \rangle.
\end{equation}
After restricting the indices such that $l_1 \le l_2 \le l_3$ and $t_1 \le t_2 \le t_3$, Eq. (\ref{cosmicvariance}) is evalulated 
using Wick's Theorem in terms of $C^T_l$, $C^E_l$ or $C^X_l$, which are the temperature, E-mode polarization and cross 
correlation power spectra respectively. It is necessary to include permutation 
factors when some of the $l$'s are equal; we include a factor of two if two $l$'s are
equal and six when all three $l$'s are equal \cite{goldberg1}. We include instrument noise in the standard 
fashion \cite{knox} and adopt parameters values (beam width and pixel noise) that are relevant 
for WMAP and Planck \cite{Hu1}.

Once we have chosen the ordering of the multiples indices, the evaluation of Eq. (\ref{cosmicvariance}) produces 
Kronecker $\delta$'s which allow us to rewrite the variance as
\begin{equation} 
     \langle N^2 \rangle = \sum_{i,j,k} \sum_{p,q,r} \sum_{\{ l,m \}}  
     W^{i,j,k}_{\{l,m\}} W^{p,q,r}_{\{l,m\}} [\mathbf{Cov}]^{i,j,k|p,q,r}_{l_1,l_2,l_3}.
\end{equation}
Defining the $(S/N)^2$ as $\langle S \rangle^2 / \langle N^2 \rangle $ we find the optimal weights by maximimizing
this ratio: 
\begin{equation}\label{funcderiv}
  2 \frac{\delta \langle S \rangle}{\delta W} 
          = \frac{\langle S \rangle}{\langle N^2 \rangle}\frac{\delta \langle N^2 \rangle}{\delta W},
\end{equation}
where
\begin{equation}
  \frac{\delta \langle S \rangle}{\delta W} 
     = \mathcal{G}^{m_1 m_2 m_3}_{l_1 l_2 l_3} b^{i,j,k}_{l_1 l_2 l_3}
\end{equation}
and
\begin{equation}
  \frac{\delta \langle N^2 \rangle}{\delta W} = 2 \sum_{p,q,r}   
     W^{p,q,r}_{\{l,m\}} [\mathbf{Cov}]^{i,j,k|p,q,r}_{l_1,l_2,l_3}.
\end{equation}
It is clear that Eq. (\ref{funcderiv}) is satisfied when we choose
\begin{equation}
   W^{p,q,r}_{ \{l,m\} } =  \sum_{ijk} \mathcal{G}^{m_1 m_2 m_3}_{l_1 l_2 l_3}
     b^{i,j,k}_{l_1 l_2 l_3} [\mathbf{Cov}^{-1}]^{i,j,k|p,q,r}_{l_1,l_2,l_3}.
\end{equation}
Now defining the quadratic form
\begin{equation}
    (\bm{q}^T \mathbf{C}^{-1} \bm{q})_{l_1 l_2 l_3} =  \sum_{i,j,k} \sum_{p,q,r} 
    b^{i,j,k}_{l_1 l_2 l_3} [\mathbf{Cov}^{-1}]^{i,j,k|p,q,r}_{l_1,l_2,l_3} b^{p,q,r}_{l_1 l_2 l_3},
\end{equation}
where $\mathbf{q}$ is a vector that contains all the possible bispectra,
and using the summation properties of the Wigner 3j symbols we find the formula for the $S/N$ when we
optimally include both temperature and polarization information about the observed CMB,
\begin{equation}\label{sndef}
  (\frac{S}{N})^2 = \sum_{2 \le l_1 \le l_2 \le l_3 \le l_{max}} 
                   \frac{(2l_1 + 1)(2l_2+1)(2l_3+1)}{4\pi}   
         \left(\begin{array}{ccc} l_1 & l_2 & l_3 \\ 0 & 0 & 0 \end{array}\right)^2
         (\bm{q}^T \mathbf{C}^{-1} \bm{q})_{l_1 l_2 l_3}.
\end{equation} 

This is a straightforward generalization of the formula in \cite{komatsu1}, where the quadratic
form in Eq. (\ref{sndef}) is simply replaced by
$b^2_{l_1 l_2 l_3}/C_{l_1}C_{l_2}C_{l_3}$.

\subsection{Reduced Bispectrum}

We follow the notation of \cite{komatsu1} in our calculation of polarization and cross correlation
reduced bispectra. Working to linear order in radiative transfer, the CMB temperature and polarization
fluctuations can be expressed as
\begin{equation}
   a^{j}_{lm} = 4\pi i^l \int \frac{d^3 \mathbf{k}}{(2\pi)^3}\Delta^{j}_l(k) Y^{*}_{lm}(\hat{\mathbf{k}}) \Phi(\mathbf{k}),
\end{equation}  
where $j = T,E$ corresponds respectively to either temperature or polarization, $\Phi(\mathbf{k})$ is the
primordial curvature fluctuation and $\Delta^{i}_l(k)$ is the 
radiation transfer function calculated by CMBFAST. We include the relevant factor of $\sqrt{\frac{(l+2)!}{(l-2)!}}$ 
in the polarization transfer function.

The reduced bispectrum, defined in Eq. (\ref{reducedbispect}), can be expressed as
\begin{equation}\label{sndef2}
   b^{i,j,k}_{l_1 l_2 l_3} = 2 f_{NL} \int r^2 dr [\beta^{i}_{l_1}(r) \beta^{j}_{l_2}(r) \alpha^{k}_{l_3}(r) 
      + \beta^{i}_{l_1}(r) \alpha^{j}_{l_2}(r) \beta^{k}_{l_3}(r)  
      + \alpha^{i}_{l_1}(r) \beta^{j}_{l_2}(r) \beta^{k}_{l_3}(r)],
\end{equation}
where
\begin{equation}
   \beta^{i}_{l}(r) = \frac{2}{\pi} \int k^2 dk P(k) j_l(kr) \Delta^{i}_{l}(k),
\end{equation}
and 
\begin{equation}
   \alpha^{i}_{l}(r) = \frac{2}{\pi} \int k^2 dk j_l(kr) \Delta^{i}_{l}(k),
\end{equation}
again where $i = T$ or $E$.
In our notation $\beta_{l}(r), \alpha_{l}(r)$ is respectively the equivalent of $b^{lin}_l(r), b^{non}_l(r)$
in the notation of \cite{komatsu1}. Defining $\tau_O$ as the present day value of conformal time, 
$\tau_R$ as the value at decoupling and $r_D = \tau_0 - \tau_R$ as the comoving distance to
the surface of last scattering, 
the region of integration for $r$ is over the sound horizon (from $\tau_O$ to $\tau_0 - 2\tau_R$).

 \begin{figure}
     \centering
     \includegraphics[width = 7.5cm, height = 7.5cm]{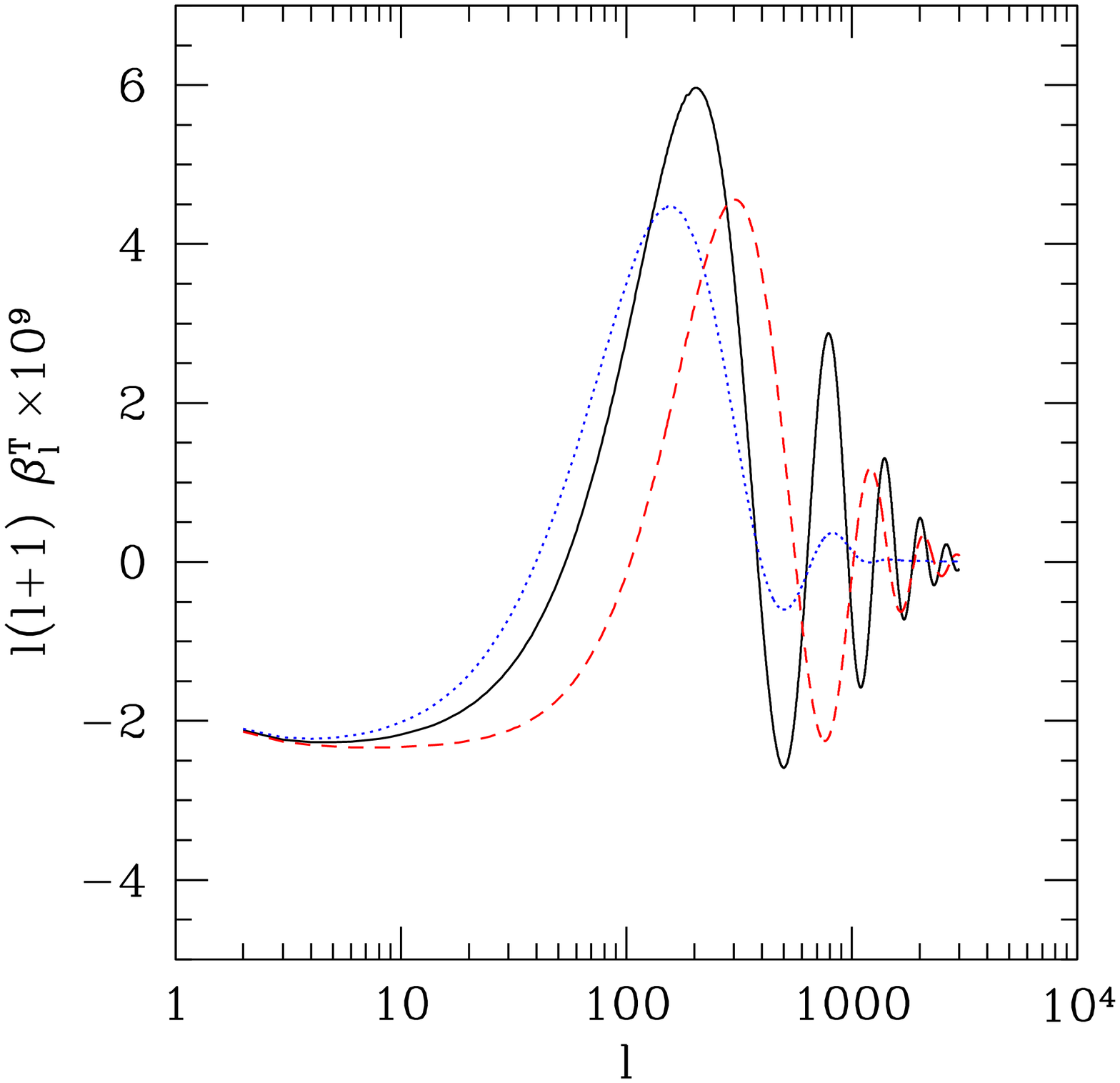}%
     \hspace{5mm}%
     \includegraphics[width = 7.5cm, height = 7.5cm]{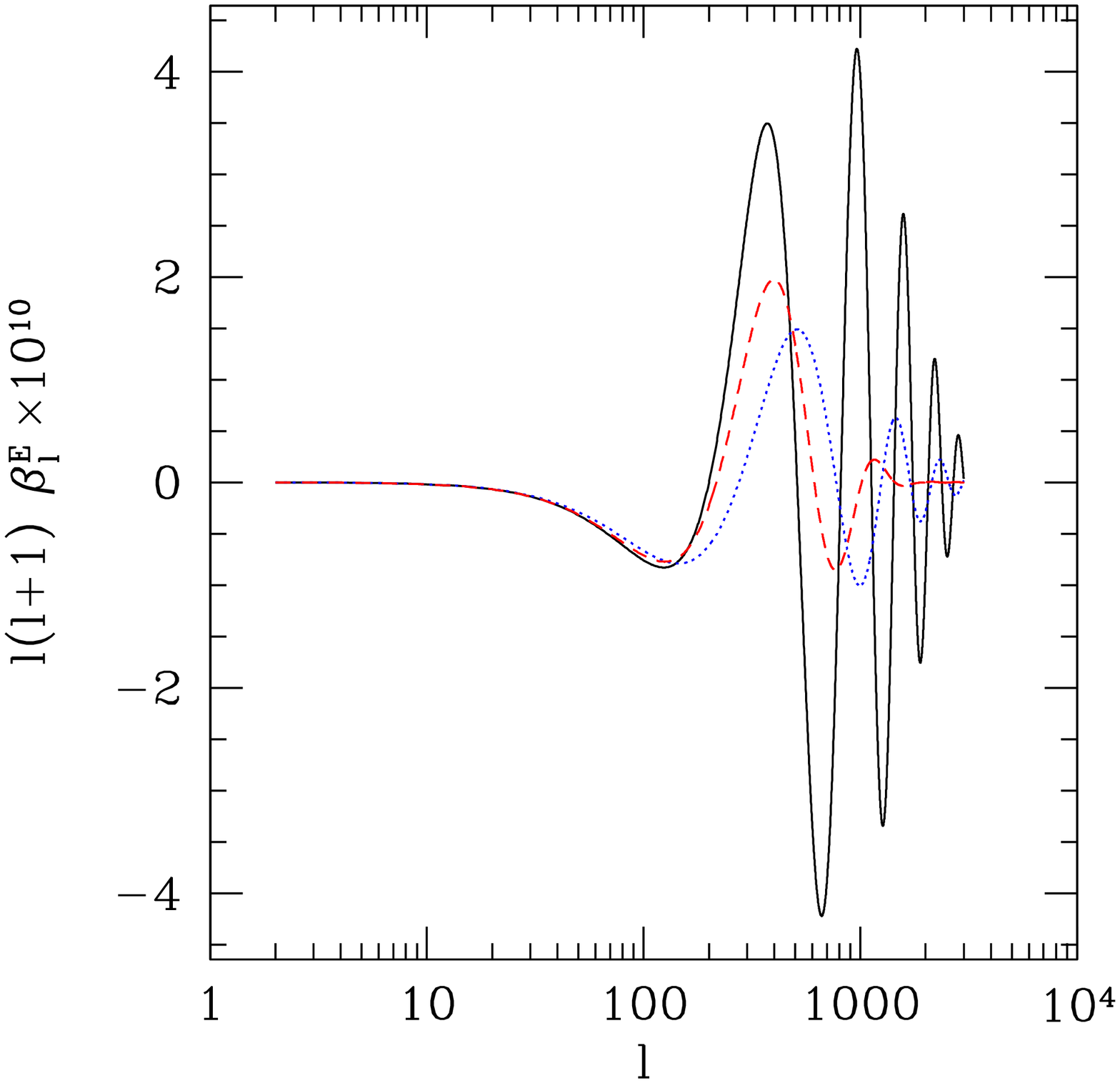}\\
     \includegraphics[width = 7.5cm, height = 7.5cm]{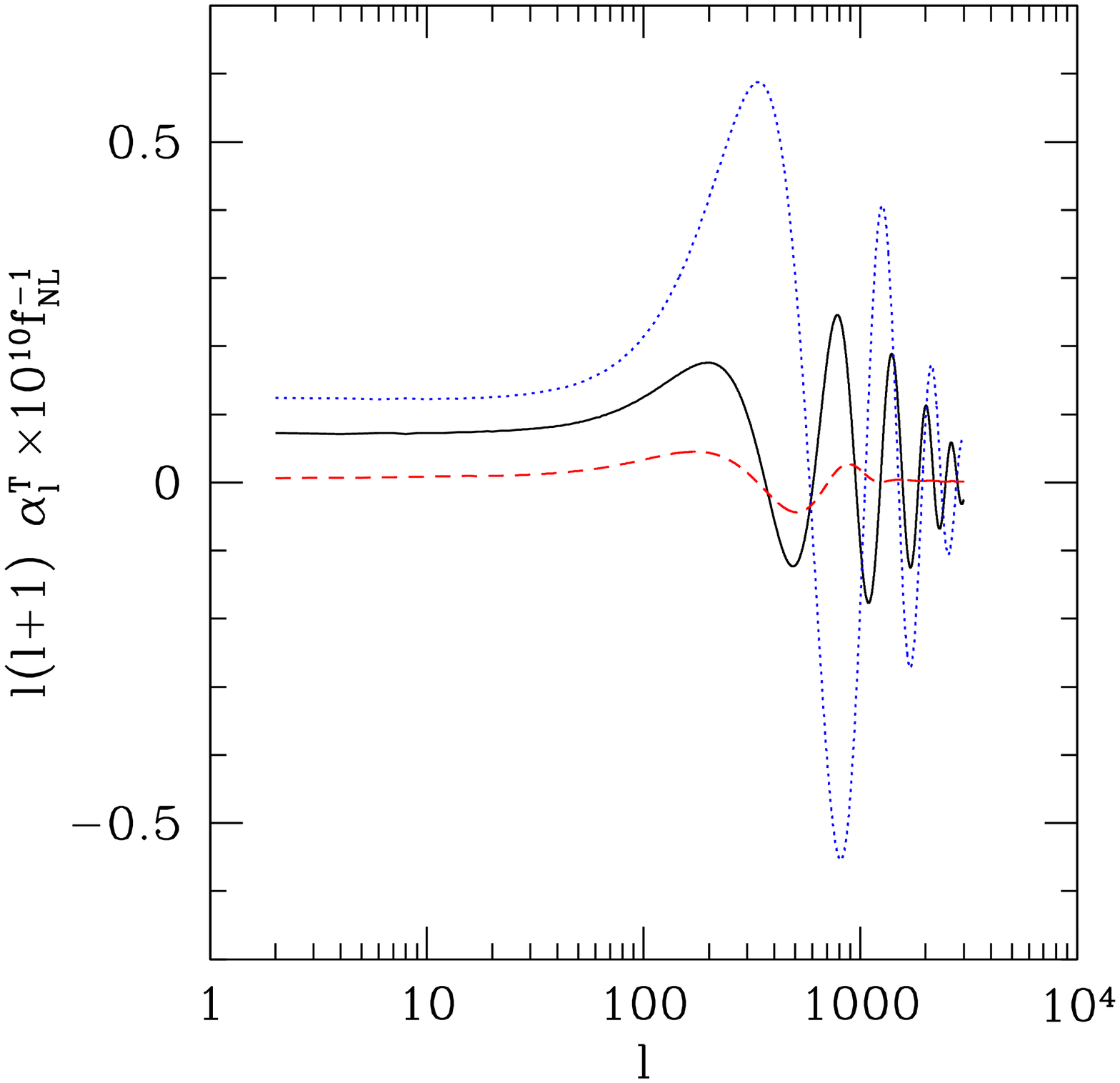}%
     \hspace{5mm}%
     \includegraphics[width = 7.5cm, height = 7.5cm]{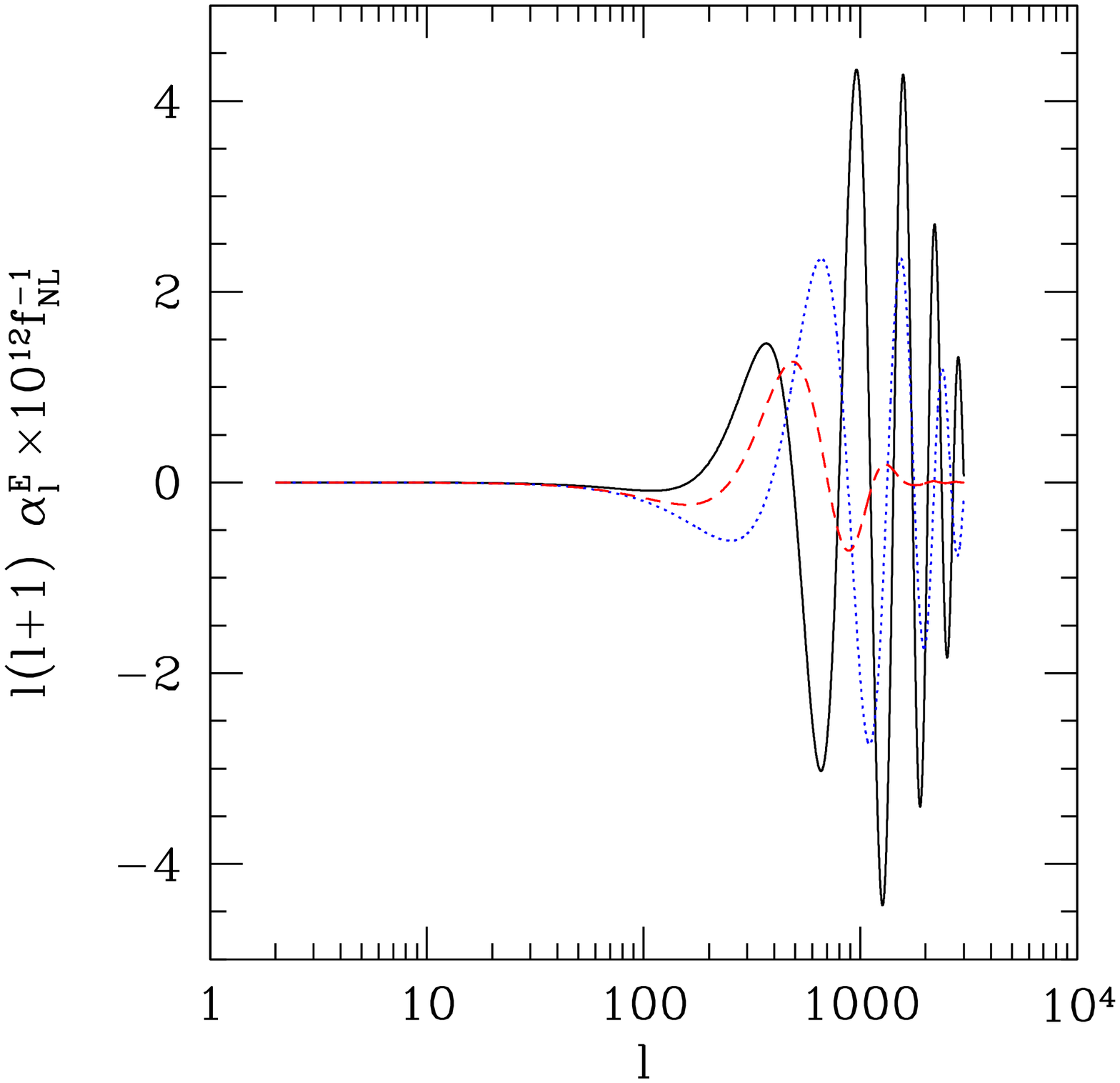}
     \caption{\label{blin} Upper Left: A plot of $\beta^T_l(r)$ vs. $l$ for values of 
      $r = \tau_0 - \tau_R$ (black, solid line), $r = \tau_0 - 0.6\tau_R$ (red, dashed line) 
      and $r = \tau_0 - 1.4\tau_R$ (blue, dotted line);
      Upper Right: same but for $\beta^E_l(r)$; 
      Lower Left: same, but for $\alpha^T_l(r)$; Lower Right: same, but for $\alpha^E_l(r)$.}	
 \end{figure}

  In Fig. \ref{blin} we see features familiar from the CMB temperature anisotropy and polarization power
  spectra. The upper level displays the temperature and polarization $\beta_l(r)$'s. On large scales
  $\beta^T_l(r)$ approaches a constant value independent of the radial distance within the sound horizon.
  The dominant mechanism on these scales is the Sachs-Wolfe effect so the radiation transfer function
  approaches $\Delta^T_l(k) = -j_l(kr_D)/3$. 
  Since there is no polarization Sachs-Wolfe effect, we see $\beta^E_l(r) \rightarrow 0$ on large scales. 
  As we approach smaller scales we see the analogue of the familiar acoustic oscillations. 

\subsection{Numerical Results}
\begin{figure}
    \centering
     \includegraphics[width = 7.5cm, height = 7.5cm]{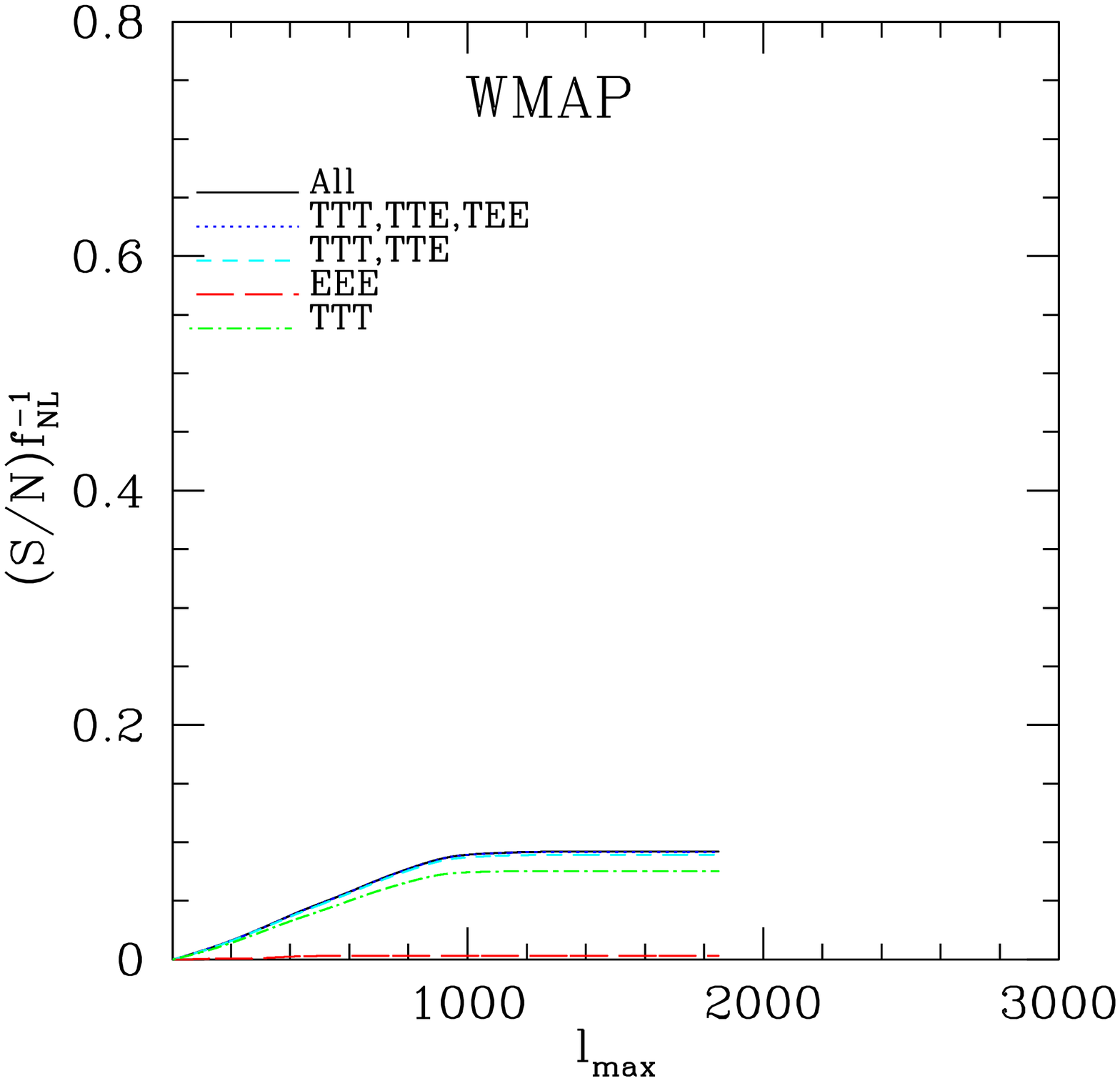}%
     \hspace{5mm}%
     \includegraphics[width = 7.5cm, height = 7.5cm]{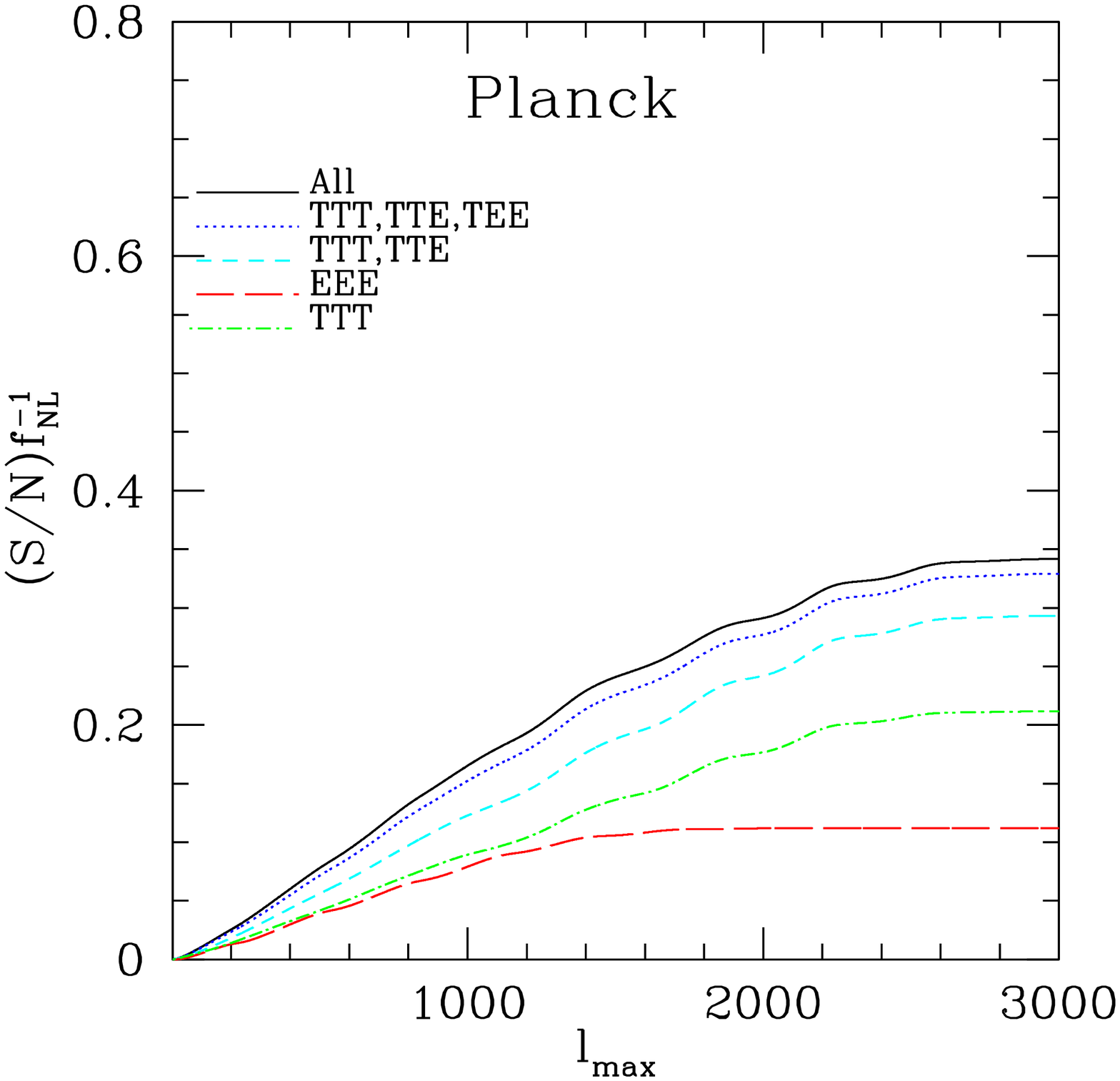}
     \caption{\label{sngl1} All figures are $(S/N)f^{-1}_{NL}$ vs. $l_{max}$ excluding 
        the effects of gravitational lensing for TTT (dot dashed green), 
        EEE (dashed red), TTT+TTE (dashed light blue), TTT,TTE+TEE (dotted blue) and all bispectra (solid black). 
        Left: WMAP; Right: Planck.}
   \end{figure}

\begin{figure}
    \centering
     \includegraphics[width = 7.5cm, height = 7.5cm]{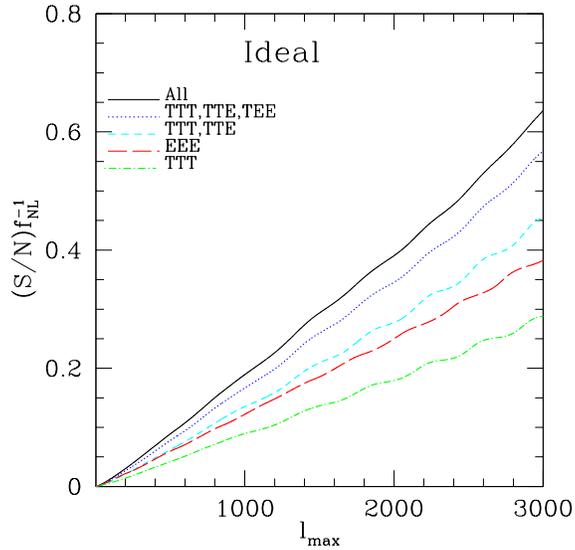}
     \caption{\label{sngl2}  Plotted is $(S/N)f^{-1}_{NL}$ vs. $l_{max}$ for an ideal experiment
       (no instrument noise and infinitesimal beam size) without the effects of gravitational lensing
       for TTT (dot dashed green), EEE (dashed red), 
       TTT+TTE (dashed light blue), TTT,TTE+TEE (dotted blue) and all bispectra (solid black).}
   \end{figure}

The curve of $S/N$ vs. $l_{max}$ will give an estimate of the minimum $f_{NL}$ measurable when both temperature
and polarization information, as well as their cross-correlations, are included in the analysis. The $S/N$
is linear in $f_{NL}$, so we can determine the minimum statistically observable $f_{NL}$ by requiring $S/N = 1$.
The CMB is only partially polarized so in experiments with relatively large noise it should be easier to measure bispectra
with fewer $E$'s than those involving three $E$s. For example WMAP will measure $TTE$ much better then $EEE$. 
Therefore it is practical to calculate the change in the constraint on $f_{NL}$ as we include bispectra with one 
additional $E$. In Fig. \ref{sngl1} the $S/N$ is forecasted for WMAP, Planck excluding the effects of gravitational lensing
and in Fig. \ref{sngl2} for an ideal experiment (no instrument noise, infinitesimal beam width) also excluding 
the effects of gravitational lensing.

Using only temperature information WMAP will be able to detect an 
$f_{NL}$ of 13.3 and Planck an $f_{NL}$ of 4.7, thus approximately recovering 
the results of the previous analysis \cite{komatsu1}. 
Our results are summarized in Table \ref{info}.
\begin{table}
\begin{center}
\begin{tabular}{c|c|c|c|c|c}
\hline
\hline
Experiment & TTT & EEE & TTT,TTE & TTT,TTE,TEE & All \\
\hline
WMAP & 13.3 & 314 & 11.2 & 10.9 & 10.9 \\
Planck & 4.7 & 8.9 & 3.4 & 3.0 & 2.9 \\
Ideal & 3.5 & 2.6 & 2.2 & 1.8 & 1.6  \\
\hline
\hline
\end{tabular}\\
\end{center}
\caption{\label{info} Minimum values $f_{NL}$ detectable with signal to noise ratio of one using stated 
bispectra with WMAP, Planck, and an Ideal experiment.}
\end{table}
When the CMB polarization is measured with good sensitivity, it appears that the inclusion of 
polarization roughly increases the sensitive of the experiments by a factor of 2.
For an Ideal experiment the minimum $f_{NL}$ is lowered to $1.6$,
which is close to the predicted size of the corrections due to second order
corrections to gravitational and hydrodynamical evolution of the CMB.

In Fig. \ref{sngl2} we see the $S/N$ curve continues to rise with a constant slope for the case 
of an ideal experiment. One might wonder why the physical scales in the radiation 
transfer functions, like the sound horizon or Silk length at the surface of last scattering, 
do not strongly influence the slope of the $S/N$ curve as $l_{max}$ increases. 
The Silk length is defined as the average distance a photon random walks before recombination. 
This signals a break down of tight coupling between the baryons and the photons that
effectively smooths out the fluctuations on small scales. One could worry that this effectively introduces an average that through the central limit theorem would reduce the level of non-Gaussianity of the resulting anisotropies, however the linear growth of 
$S/N$ with $l_{max}$ in Fig. \ref{sngl2} contradicts this intuition. In section III we will use toy models in an attempt 
to better understand our unexpected results.

\subsection{Gravitational Lensing Corrections}

   As we mentioned in the Introduction, the variance of the bispectrum estimator is the six-point function
   which includes contributions from connected two, four and six-point functions produced 
   by gravitational lensing. It is straightforward to include the influence of the two-point, 
   as the unlensed CMB power spectra are simply replaced with their gravitationally lensed counterparts in the expressions 
   for the variance of the estimator.
   This approach has been taken in \cite{komatsu1}. 
   The four-point function can become large on small scales \cite{MZ,Hu2}, so it is important to check how our 
   results change once these contributions are included. In this subsection we will investigate
   the effects of the gravitational lensing connected four-point function on the $S/N$ of the estimator we
   have defined. Since the creation of a connected four-point function is the same order in the lensing 
   potential expansion as the creation of small scale power in the two-point function we will the treat the
   two effects together in this subsection.

   In principle we could also include the effects of the connected six-point function induced by gravitational
   lensing. The leading order (in the projected lensing potential) contribution to the connected 
   six-point function is fourth order in the lensing potential and thus it should be subleading. In addition, 
   to this order in the lensing potential expansion there are extra terms in the connected two and four-point
   functions that should be included. Here we will ignore these terms and focus on those second order in 
   the lensing potential.

   We derived the optimal weights assuming the
   six-point function could be evaluated solely in terms of permutations of two-point functions. 
   Once we include the contribution from the connected four-point function the variance of our estimator 
   will increase and its $S/N$ will decrease. In this subsection we will determine the size
   of this effect. Similar work has been published on the effects of graviational lensing induced
   non-Gaussianity on the analysis of the B-mode polarization power spectrum \cite{Smith}. There it was
   shown that the non-Gaussianity reduces the information contained in the B-mode power spectrum.
 
   For simplicity we will only include CMB temperature information, but the inclusion of $E$ polarization
   is straightforward. Also we will work in the flat-sky approximation which is excellent for the relevant
   small scales. The relationship between the all-sky and flat-sky formalisms is well understood \cite{Hu3}.
   Here we adopt the conventions 
   \begin{eqnarray}\label{conventions}
     \langle a(\bm{l}_1) a(\bm{l}_2) \rangle = (2 \pi)^2 \delta^{(2)}(\bm{l}_1+\bm{l}_2) C(l_1), \\
     \langle a(\bm{l}_1) a(\bm{l}_2) a(\bm{l}_3) \rangle = (2\pi)^2 \delta^{(2)}(\bm{l}_1+\bm{l}_2+\bm{l}_3) 
    B(l_1,l_2,l_3).
   \end{eqnarray}
   In the flat-sky formalism our estimator of the three point signal is defined as
   \begin{equation}
	\hat{S} = \int \frac{d^2\mathbf{l}_1}{(2\pi)^2}\frac{d^2\mathbf{l}_2}{(2\pi)^2}
	\frac{d^2\mathbf{l}_3}{(2\pi)^2} W(\mathbf{l}_1,\mathbf{l}_2,\mathbf{l}_3)
	a(\mathbf{l}_1)a(\mathbf{l}_2)a(\mathbf{l}_3),
   \end{equation}	
   where $W(\mathbf{l}_1,\mathbf{l}_2,\mathbf{l}_3)$ is the weight function.

   Using the above conventions the expectation value of the estimator is
   \begin{equation}
	\langle S \rangle = \int \frac{d^2\mathbf{l}_1}{(2\pi)^2}\frac{d^2\mathbf{l}_2}{(2\pi)^2}
	\frac{d^2\mathbf{l}_3}{(2\pi)^2} W(\mathbf{l}_1,\mathbf{l}_2,\mathbf{l}_3)
	(2\pi)^2 \delta^{(2)}(\mathbf{l}_1+\mathbf{l}_2+\mathbf{l}_3) B(\mathbf{l}_1, \mathbf{l}_2,\mathbf{l}_3),
   \end{equation}
   and its variance is
   \begin{eqnarray}\label{flatvariance1}
   \langle N^2 \rangle &=& \int \frac{d^2\mathbf{l}_1}{(2\pi)^2}\frac{d^2\mathbf{l}_2}{(2\pi)^2}
	\frac{d^2\mathbf{l}_3}{(2\pi)^2} \frac{d^2\mathbf{l}_1'}{(2\pi)^2}\frac{d^2\mathbf{l}_2'}{(2\pi)^2}
	\frac{d^2\mathbf{l}_3'}{(2\pi)^2} W(\mathbf{l}_1,\mathbf{l}_2,\mathbf{l}_3)
	W(\mathbf{l}_1',\mathbf{l}_2',\mathbf{l}_3') \nonumber \\ 
	&&\times 
       \langle a(\mathbf{l}_1)a(\mathbf{l}_2)a(\mathbf{l}_3)a(\mathbf{l}_1')a(\mathbf{l}_2')a(\mathbf{l}_3') \rangle.
    \end{eqnarray}
   In the weakly non-Gaussian regime, meaning that the six-point function can be expressed in terms 
   of permutations of two-point functions, the estimator variance becomes
   \begin{eqnarray}\label{flatvariance}
   \langle N^2 \rangle &=& \int \frac{d^2\mathbf{l}_1}{(2\pi)^2}\frac{d^2\mathbf{l}_2}{(2\pi)^2}
	\frac{d^2\mathbf{l}_3}{(2\pi)^2} W^2(\mathbf{l}_1,\mathbf{l}_2,\mathbf{l}_3) 6 C(l_1)C(l_2)C(l_3).
   \end{eqnarray}
   This assumes that there is no strong source of non-Gaussianity. While the primordial non-Gaussianity
   is rather small there is the possibility that secondary anisotropies may cause significant non-Gaussianity
   and must be included in Eq. (\ref{flatvariance}).

   The $(S/N)^2$, defined as $ (\frac{S}{N})^2 = {\langle S \rangle^2}/{\langle N^2 \rangle} $,
   must be maximized by choosing the appropriate weight function, $W(\mathbf{l}_1,\mathbf{l}_2,\mathbf{l}_3)$. 
   We will find the optimal weight function when in the weakly non-Gaussian limit the six-point variance
   is solely determined by the Gaussian contributions and characterize the
   reduction in the $S/N$ of the estimator once the non-Gaussianity induced by gravitational
   lensing is included. Of course we could derive the optimal estimator including the connected 
   four-point function in the variance. Indeed if we discover that the gravitational lensing four-point function
   significantly reduces the $S/N$ we should modify our weight function. 
   Maximizing the signal-to-noise ratio, in an analogous fashion to above, we find that we should 
   choose our weights such that
   \begin{equation}\label{flatskyweights}
   	W(\mathbf{l}_1,\mathbf{l}_2,\mathbf{l}_3) = (2\pi)^2 \delta^{(2)}(\mathbf{l}_1+\mathbf{l}_2+\mathbf{l}_3)
	\frac{B(l_1,l_2,l_3)}{6 C(l_1)C(l_2)C(l_3)}.
   \end{equation}
   This leads to the standard formula for the $S/N$ in the flat-sky formalism
   \begin{equation}\label{flatsndef}
	(S/N)^2 = \frac{\delta^{(2)}(0)}{(2\pi)^2} 
        \int d^2\mathbf{l}_1 d^2\mathbf{l}_2 d^2\mathbf{l}_3 
        \delta^{(2)}(\mathbf{l}_1+\mathbf{l}_2+\mathbf{l}_3) 
        \frac{B^2(l_1,l_2,l_3)}{6 C(l_1)C(l_2)C(l_3)}.
   \end{equation}

   However there are corrections to the variance in Eq. (\ref{flatvariance}) from the connected four-point
   function
   \begin{equation}\label{fourpt}
	\langle a(\mathbf{l}_1)a(\mathbf{l}_2)a(\mathbf{l}_3)a(\mathbf{l}_4) \rangle
	=  (2\pi)^2\delta^{(2)}(\mathbf{l}_1+\mathbf{l}_2+\mathbf{l}_3+\mathbf{l}_4)
        T(\mathbf{l}_1,\mathbf{l}_2,\mathbf{l}_3,\mathbf{l}_4).
   \end{equation}
   Including these terms, the correction to Eq. (\ref{flatvariance}) becomes 
   \begin{eqnarray}\label{noisecorrection}
	\delta \langle N^2 \rangle =&&\frac{9}{(2\pi)^8}\int d^2\mathbf{l}_1 d^2\mathbf{l}_2
	d^2\mathbf{l}_3 d^2\mathbf{l}_1' d^2\mathbf{l}_2' d^2\mathbf{l}_3' 
	W(\mathbf{l}_1,\mathbf{l}_2,\mathbf{l}_3)W(\mathbf{l}_1',\mathbf{l}_2',\mathbf{l}_3')
       \nonumber \\ && \times 
        \delta^{(2)}(\mathbf{l}_1+\mathbf{l}_1') C(l_1)  
        \delta^{(2)}(\mathbf{l}_2+\mathbf{l}_3+\mathbf{l}_2'+\mathbf{l}_3')
        T(\mathbf{l}_2,\mathbf{l}_3,\mathbf{l}_2',\mathbf{l}_3'),
   \end{eqnarray}
   where the factor of nine comes from the cyclic permutation symmetries.
   The connected four-point function induced by gravitational lensing has been previously 
   determined in the flat-sky formalism \cite{MZ}. 

   We will determine the effects of gravitational lensing by calculating the reduction in the estimator
   signal to noise when the weights derived by ignoring gravitational lensing are used. This means that the
   weight defined in Eq. (\ref{flatskyweights}) will always be evaluated using unlensed CMB power spectra.
   However the Gaussian and non-Gaussian six-point functions in Eqs. (\ref{flatvariance}), (\ref{noisecorrection}) 
   will be evaluated with the lensed CMB power spectra when specified. 

\begin{figure}
     \centering
     \includegraphics[width = 9cm, height = 9cm]{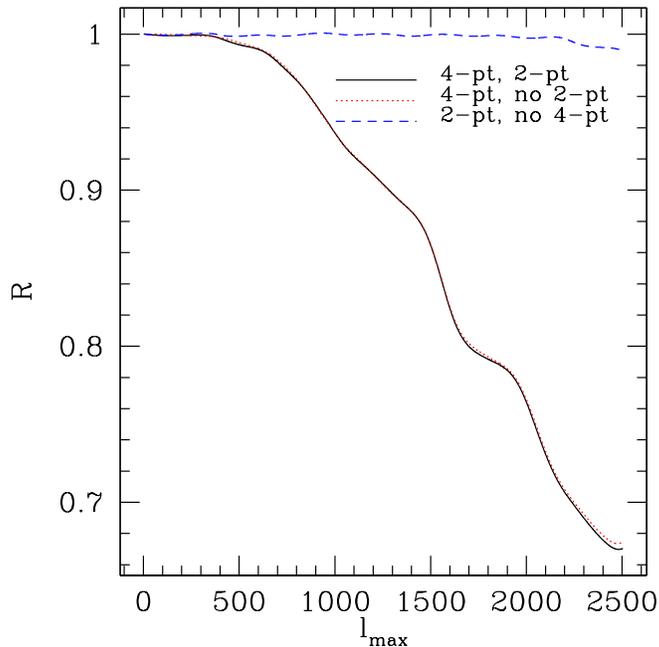}
     \caption{\label{reduc} The ratio of the $(S/N)_{GL}$ including the various forms of 
      gravitational lensing to the $(S/N)_0$ excluding the gravitational lensing is shown 
      above. The curves indicate the inclusion in the estimator variance of the 
      four-point function, but not the 
      two-point function (dotted red), both the four and two-point functions (solid black) 
      and two-point function, but not the four-point function (dashed blue).}	
   \end{figure}

   In Fig. \ref{reduc} we plot the ratio, $R = (S/N)_{GL}/(S/N)_0$, that determines the effect of 
   gravitational lensing on our ability to observe primordial non-Gaussianity. 
   We have defined the reduced $S/N$ that takes into account gravitational lensing as
   \begin{equation}
     (\frac{S}{N})^2_{GL} = \frac{\langle S \rangle^2}{\langle N^2 \rangle_{GL}},
   \end{equation}
   where $\langle N^2 \rangle_{GL}$ includes the gravitational lensing two and four-point functions as specified.
   The solid black curve indicates that the estimator variance includes both the two and four-point function
   created by gravitational lensing. While the dashed blue curve only includes the effects of the
   two-point function and the dotted red curve just the four-point function. 
   
   Figure \ref{reduc} shows that gravitational lensing does not affect the ability of WMAP
   to constrain primordial non-Gaussianity, however Planck's theoretical ability will be 
   reduced by nearly $\sim 25 \%$. The figure also shows that including the gravitational lensing correction to the two-point
   function leads only to a minor change in the signal to noise on these scales,  
   the leading effect coming from the four point function. In fact if one only includes the 
   lensing effect through the two-point function one obtains $R>1$ for some $l_{max}$ which clearly is unphysical. 
   Eventually, at high enough $l$ ($l\sim 4000)$ when the lensed CMB power spectra are significantly 
   larger than the unlensed ones, the corrections to the two point function will also decrease the signal to noise.  
   
   Recall that we assumed the CMB fluctuations were weakly 
   non-Gaussian when we derived the optimal weight functions used to construct our estimator.
   In a future work we will derive weight function including the effects of non-Gaussianity due
   to gravitational lensing \cite{babich2}.

\section{Scaling Formulae for the Signal-to-Noise Ratio}

   The shape of the $S/N$ vs. $l_{max}$ curve in Fig. \ref{sngl2}
   implies that our naive expectations of the effects of the photon diffusion are incorrect. 
   We do not see a saturation of the $S/N$ curve that we predicted earlier.
   In order to verify these counterintuitive results we will alter the CMBFAST transfer 
   functions by hand to test how sensitive the results
   are to the actual form of the transfer functions. 
   In the first case we explicitly include additional damping by multiplying the CMBFAST 
   transfer functions by an exponential $\Delta_l(k) \rightarrow \Delta_l(k) e^{-(k/k_D)^2}$, where the Silk 
   damping scale, $k_D = 500/r_D$, is chosen such that the effects appear near $l \sim 500$. In the next case
   we remove all influence of radiative transfer by choosing, $\Delta_l(k) \rightarrow \alpha j_l(kr_D)$, 
   where $\alpha$ is some constant that will always cancel in the formula for the $(S/N)^2$, Eq. (\ref{sndef}). 

   \begin{figure}
     \centering
     \includegraphics[width = 9cm, height = 9cm]{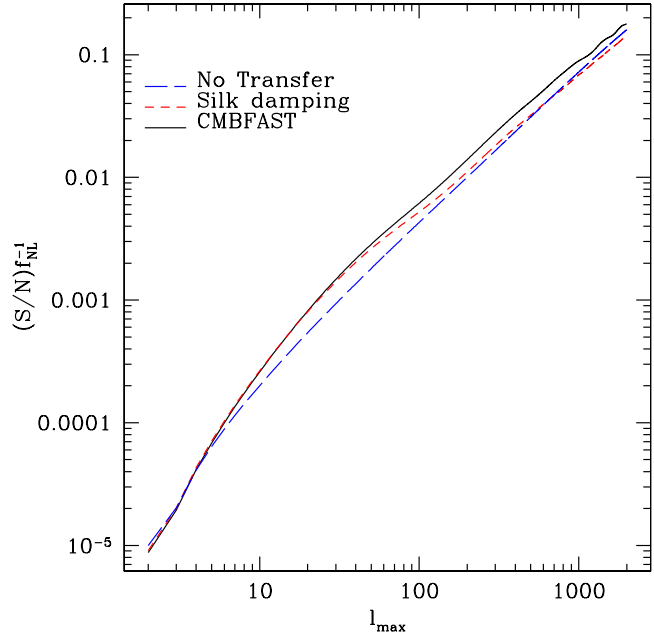}
     \caption{\label{examples} $(S/N)f^{-1}_{NL}$ vs. $l_{max}$ for TTT of the full calculation
     including the CMBFAST transfer functions (solid black), no transfer function (long dashed blue), 
     the CMBFAST transfer functions with additional damping (dashed red).}
   \end{figure}

   In Fig. \ref{examples}, all the $S/N$ curves are roughly parallel meaning that our radical altering of the transfer
   function only changes the numerical coefficients in the expressions for $S/N$. The functional dependence
   on $l_{max}$ appears to be close to the same for all three examples. This suggests that we can understand 
   these scalings by using simple toy models.

   \subsection{No Radiative Transfer}
   
    As a first example  we will calculate the $S/N$  
   in the flat sky approximation and with no radiative transfer; therefore we will simply observe the underlying
   modes restricted to the plane of the sky. 
   
    Here we adopt the following conventions 
    \begin{eqnarray}
      \langle \Phi(\bm{k}_1) \Phi(\bm{k}_2) \rangle = (2\pi)^3 \delta^{(3)}(\bm{k}_1+\bm{k}_2) P(k), \\
      \langle \Phi(\bm{k}_1) \Phi(\bm{k}_2) \Phi(\bm{k}_3) \rangle=(2\pi)^3 \delta^{(3)}(\bm{k}_1+\bm{k}_2+\bm{k}_3) 
    B(k_1,k_2,k_3),
   \end{eqnarray}
   and maintain the previously specified conventions for the CMB anisotropy Fourier coefficients, Eq. (\ref{conventions}).
   Using our assumptions of the flat sky approximation with no radiative transfer, the Fourier coefficients of the 
   temperature anisotropies can be expressed as
   \begin{equation}
      a(\bm{l}) = (2\pi)^2 \int \frac{d^3\bm{k}}{(2\pi)^3} \Phi(\bm{k}) e^{ik^zr_D} 
      \delta^{(2)}(\bm{l}-\bm{k}^{\|} r_D),    
   \end{equation}
   where $r_D$ is the distance to the surface of last scattering and $\bm{k}^{\|}$ is the Fourier wavevector parallel
   to the surface of last scattering. The power spectrum of this model is
   \begin{equation}\label{toycl}
     \frac{l^2 C(l)}{2\pi} = \frac{A}{2\pi^2}=\frac{k^3 P(k)}{2\pi^2}\equiv \Delta^2,
   \end{equation}
   where $A$ is the amplitude of the scale invariant power spectrum $P(k) = A/k^3$.
   Likewise we can find the bispectrum
   \begin{equation}
	B(l_1,l_2,l_3)  = \frac{2 f_{NL} A^2}{\pi^2} (\frac{1}{l_2^2 l_1^2} + cyc.).
   \end{equation}
   Using the standard formula $\delta^{(2)}(0) = f_{sky}/\pi$ and 
   substituting these results in Eq. (\ref{flatsndef}) gives
   \begin{equation}
     (\frac{S}{N})^2 = \frac{f_{sky} f^2_{NL} A}{6 \pi^4}
      \int d^2\bm{l}_1 d^2\bm{l}_2 d^2\bm{l}_3 
       \delta^{(2)}(\bm{l}_1+\bm{l}_2+\bm{l}_3) l^2_1 l^2_2 l^2_3 (\frac{1}{l^2_1 l^2_2} + cyc.)^2,
   \end{equation}
   and evaluating the above expression we find 
   \begin{equation}\label{toysn}
     (\frac{S}{N})^2 = \frac{4}{\pi^2} f_{sky} f^2_{NL} A l^2_{max} \ln{\frac{l_{max}}{l_{min}}}.
   \end{equation}
   The logarithm is typical of scale invariant primordial power spectra \cite{scoccimarro}. 
   If the primordial perturbations
   were generated by a Poisson process so each point in space was statistically independent, the logarithm
   would be absent and the dependence on $l_{max}$ would solely be $l^2_{max}$.
   Equation (\ref{toysn}) can be written in a more physical way by relating it to other observables,
   \begin{equation}\label{toysn1}
     (\frac{S}{N})^2 = 8 (f_{NL} \Delta)^2 N_{pix} \ln{\frac{l_{max}}{l_{min}}},
   \end{equation}  
   where $N_{pix} = f_{sky}l^2_{max}$ is the number of observed pixels. 

   \subsection{Silk Damping: Toy Model}
  
   There remains the question of why physical processes like Silk damping or cancellation due to oscillations 
   during the finite width of the last scattering surface do not cause a strong change in the slope 
   of $S/N$ curve at high $l$.
   First of all it is important to note that there are an equal number of transfer functions in the 
   numerator and denominator of Eq. (\ref{flatsndef}), so there is a sense that the effects of radiative transfer
   cancel out. Of course the transfer functions are not simple multiplicative factors that can
   be cancelled. However the saturation of $S/N$ when instrument noise or gravitational lensing comes to
   dominate the $C_l$'s, as contrasted with the continued growth of $S/N$ in the case of an ideal experiment 
   without gravitational lensing, can be understood from this perspective.
   When gravitational lensing and other secondaries or instrument noise dominates the six-point function in Eq. (\ref{flatsndef}) the
   convenient cancellation cannot occur and we can no longer recover information about primordial non-Gaussianity
   on these scales. In the case of lensing it might be possible to improve the signal to noise by using the $B$ type polarization, which
   on small scales is only generated by lensing, to constrain the deflection angle and at least partially ``unlens" the observed $T$ and $E$ fields. 

   We will attempt to explore this in the model by including the effects of Silk damping by
   introducing an exponential cutoff to mimic the effects of Silk damping on the radiation transfer
   function, 
    \begin{equation}
	a(\bm{l}) = (2 \pi)^2 \int \frac{d^3\bm{k}}{(2\pi)^3} \Phi(\bm{k}) e^{i k^z r_D} 
           \delta^{(2)}(\bm{l} - \bm{k}^{\|}r_D) e^{- k^2/2 k^2_D},
    \end{equation}
    where $k_D$ is wavevector corresponding to the Silk length below which the radiation transfer function is
    strongly damped. Repeating the above steps we find the power spectrum
    can be formally evaluated in terms of Hypergeometric U-functions as
    \begin{equation}
        C(l) =  \frac{\sqrt{\pi} A}{2 \pi l^2} e^{-l^2/l_D^2} U(1/2,0,l^2/l^2_D),
    \end{equation}
    where $l_D$ is the $2$-$d$ Fourier multiple corresponding to the Silk damping scale as $l_D = r_D k_D$.
    We can make an approximation in order to better understand the effects of Silk damping on the CMB power
    spectrum by cutting off the integral at $k \sim k_D$, then
    \begin{equation}\label{silkcl}
       C(l) = \frac{A}{\pi l^2} \frac{e^{-l^2/l^2_D}}{\sqrt{1+l^2/l^2_D}}, 
    \end{equation}
    so when $l \ll l_D$ we recover Eq. (\ref{toycl}). 
   Likewise we can evaluate the three-point functions again in order to facilitate the evaluation of 
   this integral assume that the exponentials cutoff the region of integration at $k_1, k_2 \sim k_D$.
   \begin{equation}\label{silkbispect}
      B(l_1,l_2,l_3)
         =  \frac{2 f_{NL} A^2}{\pi^2} e^{-(l^2_1+l^2_2+l^2_3)/2l^2_D} 
            [ \frac{1}{l^2_1\sqrt{1+l^2_1/l^2_D}} \frac{1}{l^2_2\sqrt{1+l^2_2/l^2_D}} + cyc.].
   \end{equation} 
   Then substituting Eq. (\ref{silkcl}) and Eq. (\ref{silkbispect}) into Eq. (\ref{flatsndef}) and assuming 
   $l \gg l_D$ 
   \begin{equation}\label{silksn}
     (\frac{S}{N})^2 = \frac{f_{sky} f^2_{NL} A l_D}{6 \pi^4} \int d^2\bm{l}_1 d^2\bm{l}_2 d^2\bm{l}_3 
	\delta^{(2)}(\bm{l}_1+\bm{l}_2+\bm{l}_3) \frac{(l_1^3 + l_2^3 + l_3^3)^2}{l_1^3 l_2^3 l_3^3}, 
   \end{equation}
   we find that leading term scales as
   \begin{equation}\label{silklimit}
	(\frac{S}{N})^2 \propto f_{sky} f^2_{NL} A l^2_{max}.
   \end{equation}
   The dependence on $l_{max}$ in Eq. (\ref{silklimit}) is nearly as strong as that in Eq. (\ref{toysn}).  
   This shows that we can still expect to recover information about $f_{NL}$ on scales where photon 
   diffusion is exponentially damping the transfer functions.  In practice, both detector noise, angular resolution 
   and secondary anisotropies will limit the smallest scale that could be used. 
   
   \subsection{Physical Arguments}

   The strongest feature of Silk damping in our toy model, the exponential damping of the CMB power spectrum 
   and the bispectrum, cancel in the expression for the $(S/N)^2$, Eq. (\ref{silksn}). However the exponential damping
   is not the only feature caused by this effect, now the power spectrum 
   Eq. (\ref{silkcl}) scales like $l^{-3}$, instead of $l^{-2}$, at large $l$. While the $(S/N)^2$ still scales
   with $N_{pix} = f_{sky} l^2_{max}$, the exact numerical coefficient that determines the slope will be reduced. 

   We can understand this behavior by considering the contribution of collapsed triangles 
   $l_1 \ll l_2 \sim l_3$. In this limit the estimator variance, ignoring factors $\mathcal{O}(1)$, is simply,
   \begin{equation}
      \sigma^2_{l_1 l_2 l_2} \sim (C_{l_1}+N_{l_1})(C_{l_2}+N_{l_2})^2,
   \end{equation}
   where $N_l$ represents both the instrument noise and any secondary anisotropy that will degrade our ability
   to recover the signal of the primary anisotropy.
   Ignoring numerical factors, the reduced bispectrum from Eq. (\ref{sndef2}) can be rewritten as 
   \begin{equation}\label{physbisp}
      b^{i,j,k}_{l_1 l_2 l_3} \sim f_{NL} \int k_1^2dk_1 k_2^2dk_2 k_3^2dk_3 
        \Delta^i_{l_1}(k_1) \Delta^j_{l_2}(k_2) \Delta^k_{l_3}(k_3) \mathcal{C}_{l_1 l_2 l_3}(k_1,k_2,k_3)
         [P(k_1)P(k_2) + cyc.],
   \end{equation} 
   where we define $\mathcal{C}_{l_1 l_2 l_3}(k_1,k_2,k_3) = \int r^2 dr j_{l_1}(k_1 r)j_{l_2}(k_2 r) j_{l_3}(k_3 r)$.
   This integral determines the geometric coupling of a triangle in Fourier space with a triangle
   on the CMB sky.

   Again in the limit of collapsed triangles, we can approximate the above coupling integral as
   \begin{equation}
      \mathcal{C}_{l_1 l_2 l_2}(k_1,k_2,k_3)  \sim j_{l_1}(k_1 r_D) \int r^2 dr j_{l_2}(k_2 r) j_{l_2}(k_3 r),
   \end{equation}
   and using the definition of the $\delta$-function,
   \begin{equation}
      \mathcal{C}_{l_1 l_2 l_2}(k_1,k_2,k_3)  \sim j_{l_1}(k_1 r_D) \frac{\delta(k_2 - k_3)}{k^2_2}.
   \end{equation}
   Here the slowly varying spherical Bessel function is evaluated at the surface of last scattering since the
   transfer function $\Delta^i_{l_1}(k_1)$ in Eq. (\ref{physbisp}) is peaked there. 

   Substituting this result into Eq. (\ref{physbisp}) we find
   \begin{equation}
       b^{i,j,k}_{l_1 l_2 l_2} \sim f_{NL} \int k_1^2 dk_1 k_2^2 dk_2 j_{l_1}(k_1 r_D) \Delta^i_{l_1}(k_1) 
       \Delta^j_{l_2}(k_2) \Delta^k_{l_2}(k_2) [P(k_1)P(k_2) + cyc.].
   \end{equation}      
   This can be evaluated as
   \begin{equation}
       b^{i,j,k}_{l_1 l_2 l_2} \sim f_{NL} C^{T,i}_{l_1} C^{j,k}_{l_2},
   \end{equation}
   where $C^{j,k}_l$ is $C^T_l, C^X_l$ or $C^E_l$ depending on the values of $j$ and $k$. 

   Now the $(S/N)^2$ for a given collapsed triangle is
   \begin{equation}\label{collapsedsn}
      \frac{b^2_{l_1 l_2 l_2}}{\sigma^2_{l_1 l_2 l_2}} \sim \frac{f^2_{NL} 
       (C^{T,i}_{l_1} C^{j,k}_{l_2})^2}{(C^i_{l_1}+N^i_{l_1})(C^j_{l_2}+N^j_{l_2})(C^k_{l_2}+N^k_{l_2})}. 
   \end{equation}  
   As long we can measure the appropriate temperature or polarization fluctuations on the scale $l_2$,
   $(C^{j,k}_{l_2})^2/(C^j_{l_2}+N^j_{l_2})(C^k_{l_2}+N^k_{l_2}) \sim 1$ and Eq. (\ref{collapsedsn}) 
   becomes independent of $l_2$.
   After integrating $l_2$ up to $l_{max}$, while keeping $l_1 \ll l_2$, we find 
   \begin{equation}
     (\frac{S}{N})^2 \propto f^2_{NL} C^T_{l_1} (r^i_{l_1})^2 l^2_{max},
   \end{equation}
   where 
   \begin{equation}
      r_{l}^i = \frac{C^{T,i}_{l}}{\sqrt{C^T_{l}C^i_{l}}}
   \end{equation}
   is the cross-correlation coefficient and again we assumed that we could resolve the primary 
   CMB anisotropies on scale $l_1$.
   
   If $i = T$, then on the very large scales we are considering $r^T = 1$ and we recover 
   \begin{equation}
     (\frac{S}{N})^2 \propto f^2_{NL} A l^2_{max},
   \end{equation}
   the result we found for our toy model, Eq. (\ref{toysn}).
   If $i = E$, then $r^E \sim 0.9$ on large scales for models without reionization and $r^E \sim 0.5$ for models 
   with a significant reionization optical depth \cite{deOliveira-Costa:2002ng}, so our conclusion still holds.

   This explains the results we found in our Silk damping toy model, Eq. (\ref{silklimit}). It is important to 
   remember that we can only detect the primordial non-Gaussianity when we can resolve the primordial anisotropies. 
   If the observed CMB power spectrum for the $l$ modes from some collapse triangle is dominated by instrument 
   noise or power induced by gravitational lensing, then cancellation of the Silk damped power spectra in 
   Eq. (\ref{collapsedsn}) will not occur and we will observe the type of saturation seen in Fig. \ref{sngl1}. 

   \section{Conclusion}

   The wealth of recent observational data has allowed the kinematics of the standard cosmological 
   model to be rigorously tested, now we must turn to the theory of its initial conditions. 
   Slow-roll inflation has become the standard scenario used to explain the initial conditions of cosmology.
   While slow-roll inflation makes several predictions,
   the Gaussianity of the underlying curvature fluctuations may be the most robust and therefore should be
   tested. Since the CMB contains additional information in its polarization patterns, 
   we calculated the increase in the Signal-to-Noise ($S/N$) ratio of the optimal cubic estimator 
   when polarization information is included. 
   The improvement in WMAP is small, from $f_{NL} \sim 13.3$ for just temperature to $10.9$ for both T and E.
   Since WMAP is too noisy to measure $E$ with high S/N we should not expect a large change. 
   For Planck the improvement is from $f_{NL} \sim 4.7$ to $2.9$ and we find
   that an Ideal experiment, with no instrument noise and infinitesimal beam width, 
   improves from $f_{NL} \sim 3.5$ to
   $1.6$. With an Ideal experiment cosmic variance limited up to $l \sim 3000$ it might
   be possible to observe the three-point function produced by non-linearites in General Relativity. 
    
   We also explored how the four-point function induced by gravitational lensing would degrade our
   estimator's $S/N$. For WMAP there is very little effect, however the constraints from Planck can
   be reduced by $25 \%$. For the next generation of CMB experiments which will measure $l > 1000$
   with good sensitivity, the estimator used to constrain the primordial non-Gaussianity should
   be derived including the effects of gravitational lensing. 

   The scaling with $l_{max}$ was shown to be related to the total number of observed independent pixels
   on the sky, implying that $(S/N)^2 \propto f_{sky} l^2_{max}\ln{(l_{max}/l_{min})}$. If the underlying distribution
   was Poisson the logarithim would be absent implying that the total $(S/N)^2$ simply scales 
   as $N_{pix} = f_{sky}l^2_{max}$.
   The functional dependence on $l_{max}$ of our full calculation using the radiation transfer functions 
   produced by CMBFAST agreed quite well with the prediction of our toy model.  
   We showed that  Silk damping did not  reduce the signal available from small scales appreciably. 
   By using a toy model it was shown that this perhaps unexpected result is caused by 
   the contribution from collapsed triangles. 

   Our ability to constrain primordial non-Gaussianity on small scales crucially depends on our ability
   to measure the primordial anisotropies on those scales. While we focused on gravitational lensing in
   this paper, there are many other mechanisms (thermal Sunyaev-Zeldovich (SZ), kinetic SZ, 
   Ostriker-Vishiniac, etc.. ) that produce additional Gaussian and non-Gaussian CMB fluctuations
   on these extremely small scales.  
   The influence of these mechanism on our estimator is difficult to determine because of the non-linear physics 
   involved and their highly non-Gaussian nature. Futher work is needed in order to understand how these effects
   will change the conclusions of this paper. These effects further strengthen the case for measuring the CMB polarization 
   on small angular scales, as the above secondaries are not expected to be significantly polarized. 

\begin{acknowledgments}   
   We would like to thank Eiichiro Komatsu, Antony Lewis and Oliver Zahn for helpful discussions.
\end{acknowledgments}


\begin{thebibliography}{36}

        \bibitem{maldacena}
	  	{J. Maldacena, JHEP 0305 (2003) 013}
		
	\bibitem{cyclic} 
	{P.J. Steinhardt and N. Turok, astro-ph/0404480 and references therein}	

	\bibitem{mukh}
	{C. Armendariz-Picon, T. Damour and V. Mukhanov, Phys. Lett. B 458 (1999) 209; J. Garriga and V. F. Mukhanov, 
	Phys. Lett. B 458 (1999) 219}

        \bibitem{creminelli}
		{P. Creminelli, JCAP 0301 (2003) 003.}
		
	\bibitem{dbi}
	M. Alishhiha, E. Silverstein and D. Tong, hep-th/0404084
		
	\bibitem{nima}
		{N. Arkani-Hamed, \textit{et.al.}, JCAP 0404 (2004) 001.}

	\bibitem{decay}
		{G. Dvali, A. Gruzinov and M. Zaldarriaga, Phys. Rev. D \textbf{69} (2003) 023505.}

	\bibitem{curvaton}
		{D.H. Lyth, C. Ungarelli and D. Wands, Phys. Rev. D \textbf{67} (2003) 023503.}

\bibitem{Creminelli:2004yq}
P.~Creminelli and M.~Zaldarriaga,
arXiv:astro-ph/0407059.

\bibitem{Gruzinov:2004jx}
A.~Gruzinov,
arXiv:astro-ph/0406129.

        \bibitem{babich1}
		{D. Babich, P. Creminelli and M. Zaldarriaga, astro-ph/0405356.}

	\bibitem{decayNG}
		{M. Zaldarriaga, Phys. Rev. D \textbf{69} (2004) 043508.}

	\bibitem{carroll}
		{T. Pyne and S. Carroll, Phys. Rev. D \textbf{53} (1996) 2920-2929.}

	\bibitem{bartolo}
		{N. Bartolo, S. Matarrese and A. Riotto, JCAP 0401 (2004) 003.}

	\bibitem{cremzal}
	{P. Creminelli and M. Zaldarriaga, astro-ph/0407059}
	 
	\bibitem{wmapurl}
		{WMAP Science Webpage: http:$\backslash\backslash$lambda.gsfc.nasa.gov}

	\bibitem{wmapfnl} 
		{E. Komatsu, \textit{et.al.}, Astrophys.J.Suppl. \textbf{148} (2003) 119-134.}

	\bibitem{komatsu1}
		{E. Komatsu and D.N. Spergel, Phys. Rev. D \textbf{63}, (2001) 063002.}

	\bibitem{MZ}
		{M. Zaldarriaga, Phys. Rev. D \textbf{62}, (2000) 063510.}

	\bibitem{Hu2}
		{W. Hu, Phys. Rev. D \textbf{64}, (2001) 083005.}

        \bibitem{seljak}
		{U. Seljak, Astrophys. J. \textbf{463}, (1996) 1.}

	\bibitem{zahn}
		{O. Zahn, D. Babich and M. Zaldarriaga, in preparation.}

	\bibitem{bunn1}
		{E.F. Bunn, Phys. Rev. D \textbf{65}, (2002) 043003.}

	\bibitem{bunn2}
		{E.F. Bunn, \textit{et.al.}, Phys. Rev. D \textbf{67}, (2003) 023501.}

	\bibitem{antony}
		{A. Lewis, Phys. Rev. D \textbf{68}, (2003) 083509.}

	\bibitem{goldberg2}
		{D.M. Goldberg and D.N. Spergel, Phys. Rev. D \textbf{59}, (1999) 103002.}
		
	\bibitem{wmapdata}
		{D.N. Spergel, \textit{et.al.}, Astrophys.J.Suppl. \textbf{148} (2003) 175.}

	\bibitem{goldberg1}
		{D.M. Goldberg and D.N. Spergel, Phys. Rev. D \textbf{59}, (1999) 103001.}

	\bibitem{knox}
  		{L. Knox, Phys. Rev. D \textbf{52}, (1995) 4307-4318.}

	\bibitem{Hu1}
		{W. Hu, Phys. Rev. D \textbf{65} (2002) 023003. }

	\bibitem{komatsu2}
		{E. Komatsu, D.N. Spergel and B.D. Wandelt, astro-ph/0305189.}

	\bibitem{dodelson}
		{W. Hu and S. Dodelson, Ann. Rev. Astron. Astrophys. \textbf{40} (2002) 171.}

	\bibitem{Smith}
		{K. Smith, W. Hu and M. Kaplinghat, astro-ph/0402442.}

	\bibitem{Hu3}
		{W. Hu, Phys. Rev. D \textbf{62}, (2000) 043007.}

	\bibitem{babich2}
		{D. Babich and M. Zaldarriaga, in preparation.}

	\bibitem{scoccimarro}
		{R. Scoccimarro, E. Sefusatti and M. Zaldarriaga, Phys. Rev. D \textbf{69}, (2004) 103513.}
	
	\bibitem{deOliveira-Costa:2002ng}
	A.~de Oliveira-Costa {\it et al.},
	Phys.\ Rev.\ D {\bf 67}, 023003 (2003).


   \end{thebibliography}
\end{document}